\documentclass[prl,twocolumn,showpacs,superscriptaddress,preprintnumbers]{revtex4}

\usepackage{graphicx}
\usepackage{dcolumn}
\usepackage{bm}
\usepackage{color} %

\begin{document}

\preprint{APS/123-QED}

\title{Superconductivity in Cu$_{x}$IrTe$_{2}$ driven by interlayer hybridization}

\author{M. Kamitani}
\email{kamitani@ce.t.u-tokyo.ac.jp}
\affiliation{Department of Applied Physics and Quantum-Phase Electronics Center (QPEC), University of Tokyo, Hongo, Tokyo 113-8656, Japan}

\author{M. S. Bahramy}%
\affiliation{Correlated Electron Research Group (CERG), RIKEN-ASI, Wako 351-0198, Japan}%

\author{R. Arita}%
\affiliation{Department of Applied Physics and Quantum-Phase Electronics Center (QPEC), University of Tokyo, Hongo, Tokyo 113-8656, Japan}%

\author{S. Seki}%
\affiliation{Department of Applied Physics and Quantum-Phase Electronics Center (QPEC), University of Tokyo, Hongo, Tokyo 113-8656, Japan}%

\author{T. Arima}%
\affiliation{Department of Advanced Materials Science, University of Tokyo, Kashiwa 277-8561, Japan}%

\author{Y. Tokura}%
\affiliation{Department of Applied Physics and Quantum-Phase Electronics Center (QPEC), University of Tokyo, Hongo, Tokyo 113-8656, Japan}%
\affiliation{Correlated Electron Research Group (CERG), RIKEN-ASI, Wako 351-0198, Japan}%
\affiliation{Cross-Correlated Materials Research Group (CMRG), RIKEN-ASI, Wako 351-0198, Japan}%

\author{S. Ishiwata}%
\affiliation{Department of Applied Physics and Quantum-Phase Electronics Center (QPEC), University of Tokyo, Hongo, Tokyo 113-8656, Japan}%

\date{\today}

\begin{abstract}
The change in the electronic structure of layered Cu$_{x}$IrTe$_{2}$ has been characterized by transport and spectroscopic measurements, combined with first-principles calculations. The Cu-intercalation suppresses the monoclinic distortion, giving rise to the stabilization of the trigonal phase with superconductivity. Thermopower and Hall resistivity measurements suggest the multiband nature with hole and electron carriers for this system, which is masked by the predominance of the hole carriers enhanced by the interlayer hybridization in the trigonal phase. Rather than the instability of Ir $d$ band, a subtle balance between the interlayer and intralayer Te-Te hybridizations is proposed as a main factor dominating the structural transition and the superconductivity. 
\end{abstract}

\maketitle

The proximity of a superconducting phase to the other quantum ordered phase provides a fertile ground for emergent electronic features. Transition metal dichalcogenides, MX$_{2}$, with layered structure have provided renewed interest for the apparently strong interplay between superconducting and charge density wave (CDW) instability, which is implicated by the trace of the CDW transition temperature falling on top of the dome-like $T_{\rm{c}}$ curve as a function of control parameter, e.g. carrier density or composition \cite{Wilson, Castro, Morosan, Morosan_PdTiSe}. IrTe$_{2}$ with Pt or Pd doping is one such new example but has an additional feature in terms of spin-orbit interaction (SOI) \cite{Pyon, Cheong}. Materials with the strong SOI have received growing attention for novel helical spin textures in  reciprocal space or real space found in topological insulators \cite{Hasan} and chiral magnets \cite{Muhlbauer, Yu}, respectively. Besides, superconductivity realized in topological insulators with spin non-degenerate Fermi surface is expected to have an unconventional gap symmetry. 

\begin{figure}[]
\includegraphics[keepaspectratio,width=8 cm]{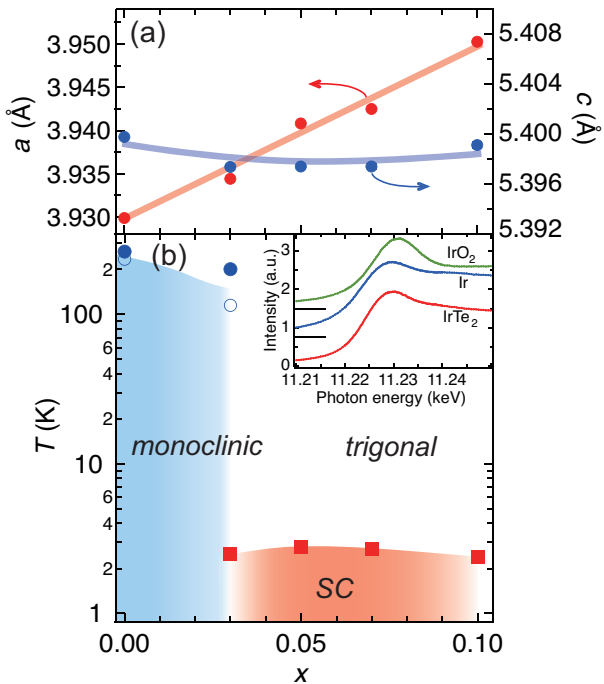}
\caption{\label{fig1} (Color online) (a) In-plane and out-of-plane lattice constants and (b) electronic phase diagram of Cu$_{x}$IrTe$_{2}$ as a function of \textit{x}. Solid lines in (a) are the guide to the eyes. Filled circles and open circles in (b) indicate the critical temperatures for structural phase transition on heating and cooling, respectively. Filled squares correspond to the superconducting transition temperature. The inset shows Ir \textit{L}$_{3}$-edge x-ray absorption spectra for IrO$_{2}$, Ir, and IrTe$_{2}$ taken at room temperature. The spectra for Ir and IrO$_{2}$ are shifted upward for clarity.
}
\end{figure}

IrTe$_{2}$ shows a structural phase transition from a high-temperature trigonal to a low-temperature monoclinic phase at around 250 K \cite{Matsumoto}. A recent study using a transmission electron microscope has revealed the presence of the superlattice modulation with a propagation vector of $\bm{q}$ = (1/5, 0, -1/5), assigned to a CDW modulation \cite{Cheong}. Considering the partially filled $t_{2g}$ orbitals indicated from their local density approximation (LDA) calculation, orbitally driven Peierls instability was proposed as an origin of the structural transition \cite{Cheong}. The importance of the orbital degree of freedom for the transition was also pointed out from the x-ray photoemission spectroscopy of Ir 4$f$ core level \cite{Ootsuki}. On the other hand, Fang et al. have postulated that the reduction of the kinetic energy of Te $p$ bands rather than the Peierls instability plays a key role in the structural transition \cite{Fang}. Despite the extensive studies on IrTe$_{2}$ with chemical substitutions, the origin of the structural transition and its relation to superconductivity are yet to be clarified.  

In this Letter, we report on transport measurements, x-ray absorption spectroscopy (XAS), and theoretical calculations for Cu$_{x}$IrTe$_{2}$ prepared by a low-temperature intercalation reaction. The monoclinic phase in IrTe$_{2}$ is dramatically suppressed and the superconductivity appears by the Cu intercalation. The Cu ion not only works as an n-type dopant but also affects the orbital hybridizations in IrTe$_{2}$. The density functional theory calculations combined with experimental results have revealed the importance of the interlayer orbital hybridization for the structural phase transition. 

Polycrystalline samples of Cu$_{x}$IrTe$_{2}$ were synthesized by two steps, including intercalation reaction in evacuated quartz tubes. First, IrTe$_{2}$ polycrystals were prepared from the stoichiometric mixture of Ir and Te powders. The mixture was heated at 900 $^{\circ}$C for 20 h, after which the sample pressed into pellets was annealed in the same condition. Second, IrTe$_{2}$ and Cu powders mixed in appropriate stoichiometry were pressed into pellets and heated at 300 $^{\circ}$C for 200 h with one intermediate grinding. Note that an impurity phase CuTe appears, when the Cu intercalation process was performed at higher temperatures than 400 $^{\circ}$C or a stoichiometric mixture of Cu, Ir, and Te was heated at 900 $^{\circ}$C. Therefore, the low-temperature intercalation reaction is necessary for the preparation of the pure Cu$_{x}$IrTe$_{2}$ phase. The samples were characterized by powder x-ray diffractions with Cu K$\alpha$ radiation. The Ir \textit{L}-edge XAS measurements were performed at 4C beamline in Photon Factory, Japan. The electrical resistivity and Hall resistivity were measured through a five-probe technique with Physical Property Measurement System (PPMS, Quantum Design). The thermopower was measured through a steady-state technique with PPMS, and the contribution from the voltage lead was subtracted. Magnetization measurements were carried out with a  superconducting quantum interference device (SQUID) magnetometer.
The electronic-structure calculations were carried out using the full-potential augmented plane-wave plus local orbital methods, as implemented in the WIEN2K code \cite{Wien2k}. Exchange correlation part of potential was treated using Perdew-Burke-Ernzerhof exchange-correlation functional \cite{Perdew}. The maximum modulus of reciprocal vectors $K_{\rm max}$ and muffin-tin radii of atoms $R_{\rm MT}$ were chosen such that $R_{\rm MT}K_{\rm max}=7$. The lattice parameters were taken from experiment \cite{Matsumoto} and the corresponding Brillouin zone of trigonal phase (monoclinic phase) was sampled by a 20 $\times$ 20 $\times$ 20  (8 $\times$ 15 $\times$ 10) $k$-mesh.

\begin{figure}[]
\includegraphics[keepaspectratio,width=8.5 cm]{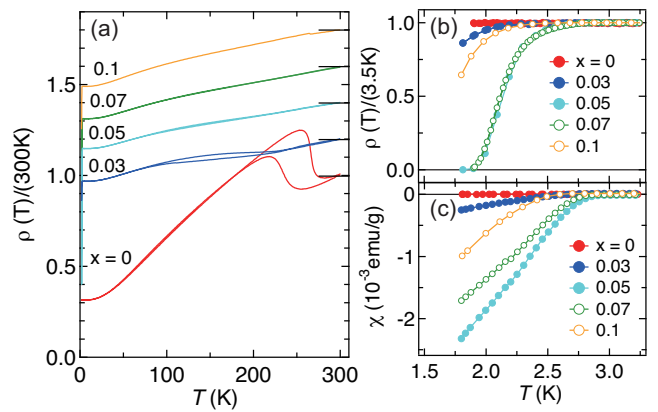}
\caption{(Color online) (a) Temperature dependence of normalized resistivity $\rho$ (T)/$\rho$ (300 K) of Cu$_{x}$IrTe$_{2}$ with $x$ ($0 \leq x \leq 0.1$). $\rho$ (T)/$\rho$ (300 K) for Cu-doped samples are shifted upwards by 0.2 for clarity. Temperature dependence of (b) $\rho$ (T)/$\rho$ (3.5 K) without external magnetic fields and (c) magnetic susceptibility under a magnetic field of 10 Oe.
}
\label{fig2}
\end{figure}

IrTe$_{2}$ crystallizes in the 1T-TaS$_{2}$ (or CdI$_{2}$) type trigonal structure ($P$-3$m$1) at room temperature \cite{Jobic}. As in the case for Cu$_{0.5}$IrTe$_{2}$, Cu$^{+}$ ions are expected to be incorporated into the octahedral site in between the Te-Te double layers \cite{CuIr2Te4}. With increasing the Cu content \textit{x}, the $a$-axis length increases monotonically, reflecting the reduction of the valence of Ir ions (see Fig. 1(a)). The change in the $a$-axis length, \textit{$\triangle$$a/a$}, with increasing $x$ from 0 to 0.1 in Pd$_{x}$IrTe$_{2}$ is less than 0.2 \% \cite{Cheong}, whereas that of Cu$_{x}$IrTe$_{2}$ is 0.5 \%, indicating that the valence of Ir ions in Cu$_{x}$IrTe$_{2}$ changes more effectively upon the chemical doping. On the other hand, the \textit{c}-axis length remains almost constant in the range of $0 \leq x \leq 0.1$, which is in contrast to the case for Cu$_{x}$TiSe$_{2}$ showing elongation of the \textit{c}-axis length by the accommodation of Cu ions \cite{Morosan}. The suppression of the elongation along the $c$-axis with increasing $x$ in Cu$_{x}$IrTe$_{2}$ can be ascribed to the strong hybridization between the Cu $s$ orbitals and the Te $p$ orbitals. The inset of Fig. 1(b) shows Ir \textit{L}$_{3}$-edge XAS for IrTe$_{2}$ together with  the references of IrO$_{2}$ and Ir, measured at room temperature. The absorption-edge energy of IrTe$_{2}$ is closer to that of Ir rather than IrO$_{2}$, suggesting that the effective valence of the Ir ions in the trigonal phase is not +4 but should be much lower. This result is consistent with the band structure calculation, as discussed later. The change in XAS spectra by the Cu intercalation is too small to be discerned (data not shown).

Figure 2(a) shows the temperature dependence of resistivity for Cu$_{x}$IrTe$_{2}$. All the samples show a metallic behavior. The anomaly with large thermal hysteresis around 200-250 K in IrTe$_{2}$ can be associated with the structural phase transition from the trigonal to monoclinic form \cite{Matsumoto}. The hump-shaped anomaly broadens and shifts to lower temperature by the Cu intercalation of $x$ = 0.03, and almost disappears for $x$ = 0.05, where the trigonal phase is stabilized down to the lowest temperature and the superconductivity with $T_{\rm{c}}$ (onset) = 2.8 K emerges, as shown in Figs. 2(b) and 2(c). Zero resistivity and clear shielding signals are observed for the samples with $x$ = 0.05 and 0.07, while for the samples with $x$ = 0.03 and 0.1 faint indications of superconductivity are found at lower temperatures.  As observed for the Pt or Pd doped IrTe$_{2}$ \cite{Pyon, Cheong}, the dome-like $T_{\rm{c}}$ vs. $x$ curve can be confirmed for Cu$_{x}$IrTe$_{2}$ in the vicinity of the phase boundary between the trigonal and monoclinic phases. 
  
\begin{figure}[]
\includegraphics[keepaspectratio,width=6.6 cm]{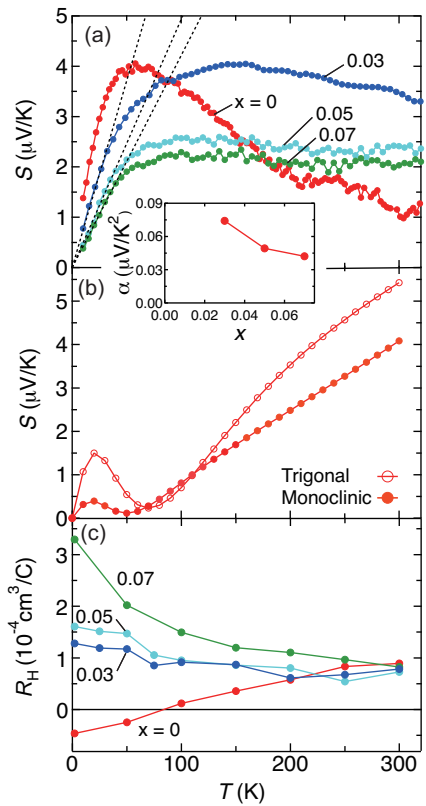}
\caption{(Color online) Temperature dependence of (a) Seebeck coefficient measured on cooling for Cu$_{x}$IrTe$_{2}$, (b) calculated Seebeck coefficient, and (c) Hall coefficient. The broken lines in the panel (a) are the fit to the data at low temperatures based on the equation $S$ = $\alpha T$. The inset shows the variation of $\alpha$ as a function of the Cu content $x$.
}
\label{fig3}
\end{figure}

Next, we show Seebeck coefficient $S$ and Hall coefficient $R_{\rm{H}}$ of Cu$_{x}$IrTe$_{2}$, both of which are less affected by the grain boundaries and thus suitable for characterizing the electronic properties of the polycrystalline samples (see Figs. 3(a) and 3(c)). The sign of $S$ is positive in all compositions and whole temperature range. This is semi-quantitatively reproduced by the density functional theory calculation, as the calculated $S$ values for the monoclinic and trigonal phases with $x$ = 0 are positive over the whole temperature (see Fig. 3(b)).

However, while $R_{\rm{H}}$ is positive for the Cu-intercalated compounds, it becomes negative below 100 K in the monoclinic phase of IrTe$_{2}$, reflecting the difference in the Fermi surfaces of both phases and the multi-band nature of the monoclinic phase. \textit{S} of IrTe$_{2}$ shows a kink at 200-250 K, which reflects the reconstruction of the Fermi surface across the structural transition, and gradually increases as decreasing temperature down to 50 K. Such a behavior in $S$ is almost smeared out for Cu$_{0.03}$IrTe$_{2}$. Here, we introduce the following expressions of $S$ and $R_{\rm{H}}$ for a two-band system with electron and hole carriers \cite{Yoshino}.
\begin{equation}
  S = \frac{n_{\rm{h}}\mu_{\rm{h}}S_{\rm{h}} +n_{\rm{e}}\mu_{\rm{e}}S_{\rm{e}}}{n_{\rm{h}}\mu_{\rm{h}}+n_{\rm{e}}\mu_{\rm{e}}} ,
\end{equation}
\begin{equation}
  R_{\rm{H}} = \frac{(n_{\rm{h}}\mu_{\rm{h}}^{2}-n_{\rm{e}}\mu_{\rm{e}}^{2})}{e(n_{\rm{h}}\mu_{\rm{h}}+n_{\rm{e}}\mu_{\rm{e}})^{2}} .
\end{equation}
Here, $n_{\rm{h}}$, $\mu_{\rm{h}}$, and $S_{\rm{h}}$ represent the carrier concentration, mobility, and Seebeck coefficient for the hole band, respectively, and $n_{\rm{e}}$, $\mu_{\rm{e}}$, and $S_{\rm{e}}$ do those for the electron band. From these equations, it is found that the sign reversal between \textit{R}$_{\rm{H}}$ and $S$ as well as the temperature-dependent sign reversal of \textit{R}$_{\rm{H}}$ in the monoclinic phase is a hallmark of the multi-band with electron and hole carriers. 

On the other hand, the multiband nature is masked in the trigonal phase by the strong interlayer hybridization giving rise to the predominance of hole conduction through the Te 5$p$ band (as seen in Fig. 4(a) and ref. \cite{Fang}). Therefore, we have performed a semi-quantitative analysis for the trigonal phase with a single-carrier model. Below 30 K, $S$ for the Cu-intercalated compounds obeys linear temperature dependence in accordance with the diffusive thermopower in a metallic system \cite{Nielsen}. With assuming the quasi-2D parabolic band \cite{Mandal}, \textit{S} can be described by the following equation: 
\begin{equation}
  S = -\frac{\pi k_{\rm{B}}^{2}}{2e\hbar^{2}d}\frac{m^{*}}{n}T ,
\end{equation}
where $2\pi\hbar$, $k_{\rm{B}}$, $d$, $n$, and $m^{*}$ represent the Planck constant, the Boltzmann constant, distance between the adjacent conduction layers, carrier concentration, and effective mass, respectively. With the use of the equation (3), $m^{*}$ for Cu$_{0.05}$IrTe$_{2}$ is calculated to be 4.3 $m_{0}$ ($m_{0}$ being the bare electron mass) by adopting the estimated carrier concentration at 2 K ($n$ = 3.9$\times$10$^{22}$ cm$^{-3}$) and the fitting parameter $\alpha$ in the formula \textit{S} = $\alpha T$. This value is much larger than that estimated from the band calculation ($m^{*}$ = 0.6 - 2.4 $m_{0}$). The discrepancy in magnitude between the experimental and the calculated $m^{*}$ as well as $S$ is likely to arise from the finite electron correlation in addition to the simplicity of the model. Nevertheless, the decrease in $\alpha$ with increasing $x$ from 0.03 to 0.07 (see the inset of Fig. 3(b)) can be ascribed to the decrease in $m^{*}$ rather than the increase in $n$, which is supported by the band calculation and the Hall coefficient data. This interpretation for $S$ qualitatively explains the change in the electronic specific-heat coefficient in the trigonal phase of Ir$_{1-x}$Pt$_{x}$Te$_{2}$ \cite{Pyon}. 

\begin{figure}[]
\includegraphics[keepaspectratio,width=7.8 cm]{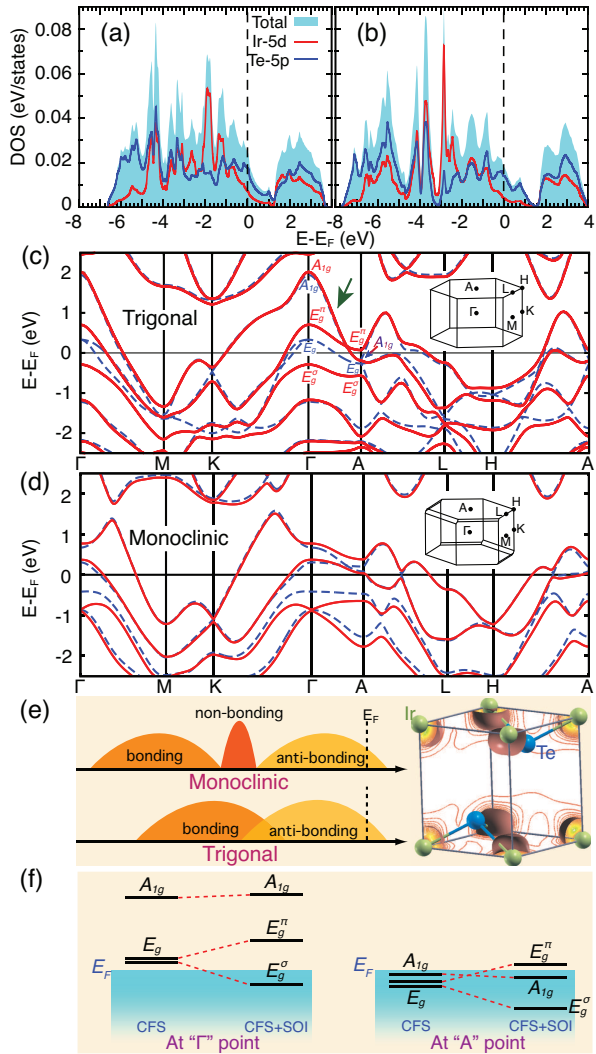}
\caption{Calculated density of states of IrTe$_{2}$ in (a) the trigonal phase and (b) the monoclinic phase. The respective band structures near the Fermi level are shown in (c) and (d). The solid (red) and dashed (blue) lines correspond to the band dispersions obtained with and without inclusion of spin-orbit interaction (SOI), respectively. (e) Relative positioning of bonding and antibonding states in each phase. The monoclinic phase has additionally nonbonding states (corresponding to the states spanning energies from -3 eV to -4.6 eV in (b)) which are mainly localized on Ir atoms. The electron density distribution of these states is shown in the inset. For the trigonal phase, the interplay between the crystal field splitting (CFS) and SOI at ƒ¡ and A points is schematically shown in (f).}
\label{fig4}
\end{figure}

Let us discuss the origin of structural phase transition in IrTe$_{2}$ and its relation to the superconductivity, based on the result of density functional theory. As shown in Figs. 4(a) and 4(b), the calculated density of states (DOS) of trigonal and monoclinic phases reveals that the Ir-$d$ states are widely spread below the Fermi level ($E_{\rm F}$) with only a minor portion of them left unoccupied. Thus, the nominal valence of Ir is neither +3 nor +4, but should be much lower. In fact, our calculated Wannier charges suggest that Ir (Te) loses (gains) only 1.1 $e$ (0.5 $e$) in the trigonal phase, which appears to be in good agreement with our XAS measurements. Due to this rather weak Ir-Te bonding, what possibly derives the phase transition in this material is the Te-Te interaction. In the high temperature trigonal phase, this interaction is much stronger between the Te atoms of neighboring layers than those within the same IrTe$_{2}$ layer. Because of this, the interlayer Te-Te distance, $d_{\rm Te}$, becomes smaller than $d_{\rm Te}$ within the same layer (3.498 \AA \ compared with 3.557 \AA \ and 3.928 \AA \ \cite{Jobic}). Consequently, the electronic bands around $E_{\rm F}$ become highly dispersive in all directions, especially along the $c$-axis (corresponding to $\Gamma$-A direction as indicated by the arrow in Fig. 4 (c)). Due to this strong interaction, the bonding and antibonding states of Te can strongly overlap with each other through Ir-$d$ states. The emergence of superconductivity can be associated with the stabilization of quasi-3D hole-like Fermi surface in the trigonal phase.

Lowering the temperature into the monoclinic phase, the bonding states move to the lower energies, allowing the system to reduce its total energy. Such an energy reduction results in the formation of a nonbonding region between the bonding and antibonding states mainly localized on Ir atoms, as shown in Fig. 4(e). As the overlap between the bonding and antibonding states has been reduced, the interlayer interaction is substantially suppressed. On the other hand, as the bonding states have moved to lower energies, the interaction of Te atoms within the same layer is enhanced; the interlayer (intralayer) $d_{\rm Te}$ increases (decreases) to 4.036 \AA \ (3.083 \AA \ and 3.812 \AA \ \cite{Jobic}). This also results in a considerable reduction in the Ir-Te bond length and hence enhances the $d$-$p$ hybridization. The overall effect of these changes is that the Ir-Ir bonds become relatively elongated along one of the in-plane axes, therefore resulting in a phase transition to the monoclinic phase. In the monoclinic phase, the interlayer interaction and hence the band dispersion along the $c$-axis ($\Gamma$-A direction) are suppressed, therefore the 2D character of Fermi surface topology becomes enhanced (see Fig.4 (d)). Such a change in the topology of FS along the $k_z$ direction due to the phase transition appears to be consistent with what has been observed in ARPES measurements \cite{ARPES}.  

At this point, it is worth briefly explaining the role of crystal filed splitting (CFS) and SOI on the electronic band dispersions of IrTe$_{2}$. As shown in Figs. 4 (c) and 4(d), the electronic states around $E_{\rm F}$ are more dramatically modified by SOI in the trigonal phase than in the monoclinic phase. Without SOI, in the former phase, due to the trigonal CFS, the predominantly $p$-type states are split into the two-fold degenerate $E_{\rm g}$ and single-fold $A_{\rm 1g}$ bands. Such a splitting substantially decreases in going from the $\Gamma$ point to A point due to the strong interlayer hybridization between Te atoms, thereby making the both bands occupied in the vicinity of A point. Turning on SOI, however, the $E_{\rm g}$ bands are split off  by nearly 1 eV into $E_{\rm g}^{\rm \sigma}$ and $E_{\rm g}^{\rm \pi}$ branches such that the former (latter) one is pushed below (above) $E_{\rm F}$. Given the strong (weak) CFS at $\Gamma$ (A) point, the overall effect of SOI appears as a change in the ordering of  $E_{\rm g}^{\rm \pi}$ and $A_{\rm 1g}$ bands along the $\Gamma$-A direction, as schematically depicted in Fig. 4 (f).  The reported topological complexity of Fermi surface in the trigonal phase \cite{Cheong} is thus expected to be due to strong interplay between the trigonal CFS and SOI.

In summary, we have experimentally and theoretically studied superconducting Cu$_{x}$IrTe$_{2}$ synthesized by low-temperature intercalation reaction. Thermopower and Hall resistivity reveal the multiband nature with electron and hole carriers in the monoclinic phase and the predominance of hole carriers in the trigonal phase stabilized by the Cu doping. Contrary to the previous reports \cite{Pyon, Cheong, Ootsuki}, the band structure calculations combined with XAS measurements suggest that the valence of Ir ions is much lower than +4, and the dramatic change in the interlayer and intralayer hybridizations plays an important role in the structural phase transition, rather than the instability of Ir $t_{\rm 2g}$ orbitals. These results imply that the suppression of the monoclinic phase and the emergence of superconductivity are not only caused by the chemical potential shift in the rigid band but by the enhancement of interlayer orbital hybridization. The highly dispersive bands along the $c$-axis, which come from the strong interlayer hybridization, are expected to play an important role in producing the superconductivity in this multiband system.

\begin{acknowledgments}
The authors thank H. Sakai for enlightening discussions. This study was in part supported by the Grant-in-Aid for Scientific Research (Grant No.23685014) from the MEXT, and by Funding Program for World-Leading Innovative R\&D on Science and Technology (FIRST Program), Japan.
\end{acknowledgments}

\end{document}